# Fault Location Estimation by Using Machine Learning Methods in Mixed Transmission Lines

Serkan Budak[1*], Bahadır Akbal[2]

[1] Department of Electrical and Electronics Engineering/Konya Technical University,
[2] Department of Electrical and Electronics Engineering/Konya Technical University,
[1*](sbudak@ktun.edu.tr) Email of the corresponding author

**Abstract**

Overhead lines are generally used for electrical energy transmission. Also, XLPE underground cable lines are generally used in the city center and the crowded areas to provide electrical safety, so high voltage underground cable lines are used together with overhead line in the transmission lines, and these lines are called as the mixed lines. The distance protection relays are used to determine the impedance based fault location according to the current and voltage magnitudes in the transmission lines. However, the fault location cannot be correctly detected in mixed transmission lines due to different characteristic impedance per unit length because the characteristic impedance of high voltage cable line is significantly different from overhead line. Thus, determinations of the fault section and location with the distance protection relays are difficult in the mixed transmission lines. In this study, 154 kV overhead transmission line and underground cable line are examined as the mixed transmission line for the distance protection relays. Phase to ground faults are created in the mixed transmission line, and overhead line section and underground cable section are simulated by using PSCAD/ EMTDC ™. The short circuit fault images are generated in the distance protection relay for the overhead transmission line and underground cable transmission line faults. The images include the R-X impedance diagram of the fault, and the R-X impedance diagram have been detected by applying image processing steps. The regression methods are used for prediction of the fault location, and the results of image processing are used as the input parameters for the training process of the regression methods. The results of regression methods are compared to select the most suitable method at the end of this study for forecasting of the fault location in transmission lines. When looking at the method and performance criteria used in the overhead transmission line fault location study, it is the Linear Regression (Robust Linear) method that gives the most accurate results with RMSE 0.017652. When looking at the method and performance criteria used in the underground cable transmission line fault location study, it is the Linear Regression (Stepwise Linear) method, which gives the most accurate results with RMSE 0.0060709. When the accuracy of the method was examined, it was seen that it was higher than other methods.

# 1. Introduction

Overhead line systems dominate today in the transmission of electrical energy. Overhead line systems are transmission lines that have proven their operational reliability and functional use. Installation costs of overhead lines are low and their useful life is higher than underground cable lines (Wedepohl & Wilcox, 1973).

Nowadays, underground cables are used in distribution lines, especially in places where there is a high density of people, taking into account the operational safety, human-environmental health and the economy of the enterprise. With the new technological developments in insulation systems in high voltage XLPE underground cables, cables have been used in transmission lines. Nowadays, electricity demand is increasing rapidly in big cities. All over the world, underground power cable installations have begun to replace some of the overhead transmission lines due to environmental factors in densely populated areas. Protection systems for power transmission lines are one of the most important parts of power systems.

Determining the fault location in transmission lines is a desired feature in the protection scheme. The increasing complexity of modern power transmission systems, fault location has increased substantially the importance of research studies in recent years. If the location of a fault is known or can be estimated with high accuracy, the fault can be rectified quickly. Elimination of the fault is of great importance as it reduces customer complaints, downtime, operating cost, loss of income and maintains the stability of the system. For this reason, studies are carried out to use mixed transmission lines (line with overhead and underground cable) with high efficiency (Sadeh & Afradi, 2009).

When the studies in the literature are examined, studies have been carried out on artificial intelligence methods by using current and voltage information in the problem of fault location in mixed transmission lines. In addition, the traveling waves method is often used.

In this study, unlike other studies, the fault location estimation was made with regression methods using the images obtained from the distance protection relay.

# 2. Material and Method

In order to protect transmission lines against short circuit faults and to provide information about the location of the fault, distance protection relays are widely used today. Since distance protection relays detect impedance-based fault and fault locations according to current and voltage magnitudes, fault location cannot be detected correctly due to different impedance per unit length in mixed transmission lines (Han & Crossley, 2013, 2015).

In this study, images taken from the R-X impedance diagram for short circuit faults occurring in the overhead transmission line and underground cable transmission line were applied to the image processing steps and a data set was created to be given to the Matlab Regression Learner application. Performance results and fault location estimates of the methods used in Regression Learner application are given in the findings section.

## 2.1. Mixed Transmission Lines

The underground cable line can replace a part of the existing overhead line in high voltage transmission lines, but the characteristic impedance of the cables is significantly different to the overhead line (Tziouvaras, 2006). The series inductance of an underground cable is 30-50% less than an overhead line, but the shunt capacity of a cable is 30-40 times greater than that of an overhead line (Tziouvaras, 2006).

Quick and accurate fault location reduces costs for locating the fault. In addition, it speeds up operations that need to be done quickly, such as fault repair and re-commissioning of the transmission line. Therefore, it will reduce production, usage and revenue losses due to interruptions. Therefore, quick and accurate estimation and determination of the fault is very important in terms of operating continuity of the system, operational safety and operating costs.

## 2.2. Image Processing Method

Image processing is a method applied with different techniques to obtain different information from images that have been digitized. In the design and analysis of image processing systems, the image is expressed mathematically. When images are digitized using image processing method, a large matrix form is obtained. In large matrix training stages, the error rate is high and the processing time is quite long. For this reason, matrix size, processing time and error rate are reduced by applying different statistical methods. This process is created by extracting the features of the images before the training is applied.

With the Gray Level Co-occurrence Matrices (GLCM) function, we can obtain different statistical properties for each image. Statistical methods used in feature extraction were used in the study. These are mean, entropy, variance, difference, contrast, inverse difference moment, energy, correlation, cluster shade, cluster prominence, sum entropy, sum mean, difference entropy, sum variance values. Properties of 1x20 size image are extracted for each image (Budak, 2020).

The training data set was created by using images image processing methods. The applied image processing steps are shown in Figure 1 (Budak, 2020).

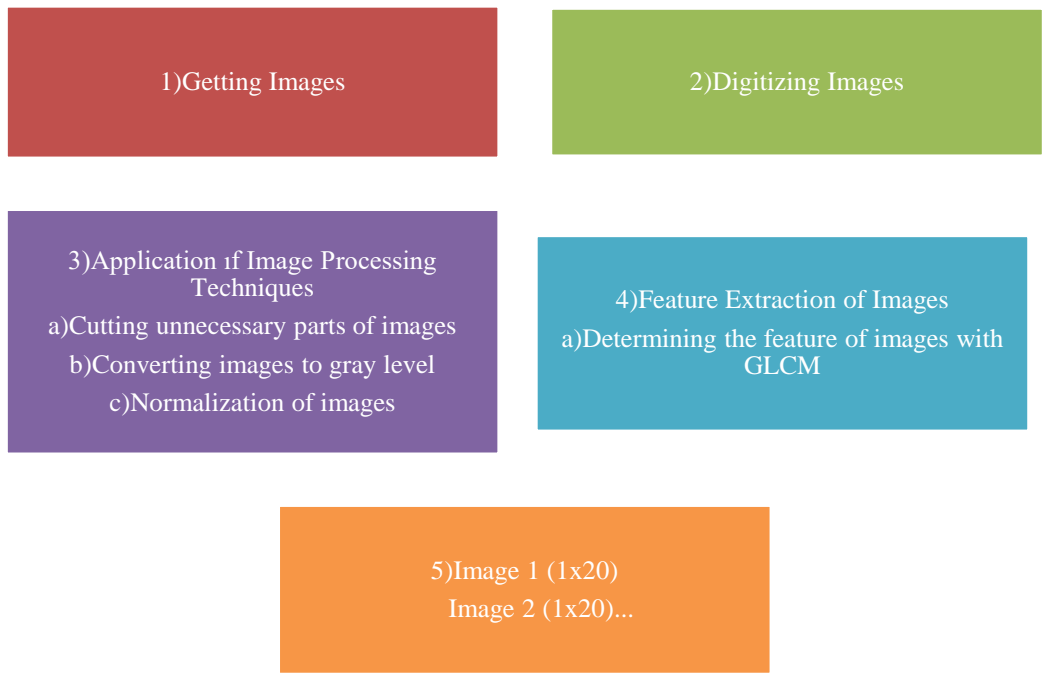

Figure 1. Applied İmage Processing Steps

## 2.3. Matlab Regression Learner App

Regression Learner is an application that provides the opportunity to compare machine learning methods widely used in the literature and determine the most suitable model (MathWorks). İnput and output data sets are given to the application, data is checked, features are selected, verification schemes are determined and performance results are obtained by training with different models. Results of each training model can be visualized, access a Response Plot, Predicted vs. Actual Plot and Residual Plot. Performance criterion used to predict training models and to satisfy them. The performance criterion mentions the model and allows us to choose the best model. In the application, the estimated training time and speed information is given to the user. Matlab Regression educational models used in the Learner application Linear Regression, Tree, Ensemble Trees, Support Vectors Machines (SVM), Gaussian Process Regression and the best educational model can automatically find their sub methods (MathWorks). It can be used by automatically training multiple models at the same time in its application. In this application, training models can be used quickly and the best training model can be determined. In this way, there is no need to write separate and different codes, and it provides the opportunity to test many training models at the same time (Özleyen, 2019).

## 2.4. Simulation Study

In order to create a mixed transmission line, two 154 kV, 50 Hz, 200 km and 50 km overhead transmission lines and a 10 km underground cable line between the overhead lines are modeled in the PSCAD ™ / EMTDC ™ simulation program. The transmission line is fed from two generator generation sources connected to two three-phase power transformers. For the overhead transmission line, 154 kV single circuit power transmission line, 1272 MCM conductor cross sections and 'PB' pole type are designed. 89/154 kV, 2XS(FL)2Y cable type is designed for underground cable transmission line. In the connections of transformers, the low voltage side has been selected as 11 kV delta and the high voltage side as 154 kV star. A fixed load of 20 MW was used in the system. In the study, the situation of short circuit faults occurring both in the overhead transmission line and in the underground cable transmission line was observed. Studies have been done by creating a phase A-ground short circuit, which is the most common single-phase ground fault. The mixed transmission line model is shown in Figure 2.

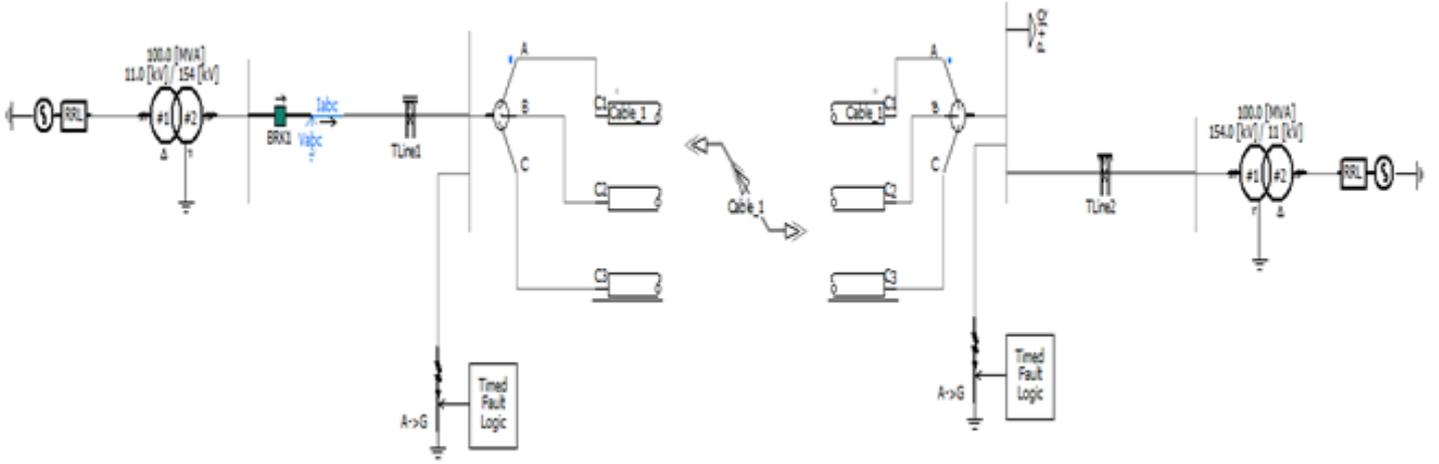

Figure. 2 Mixed Transmission Line Model

Figure 3 shows sample images occurring in phase a-ground short circuit faults formed at the 60th km of the overhead line section of the mixed transmission line and at the 5th km of the underground cable line section.

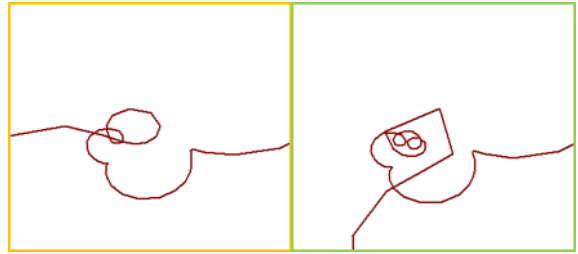

Figure. 3 Sample Images Obtained From the R-X Impedance Diagram in Mixed Transmission Lines

Root Mean Square Error (RMSE) and percentage error value were used to evaluate the results obtained in fault location estimation studies. The RMSE value is always positive and close to zero indicates the best value. Equation 1 and Equation 2 contain percent error and RMSE equations, respectively (Karasu, Altan, Saraç, & Hacıoğlu, 2018).

$$\%Error\ Value = \left|\frac{Actual\ fault\ location - Calculated\ fault\ location}{Total\ length\ of\ the\ line}\right| * 100 \quad (1)$$

$$RMSE = \sqrt{\frac{\sum_{i=1}^{n}(xi-yi)^2}{n}} \quad (2)$$

## 3. Results and Discussion

40 number phase a-ground faults were created at 5,10,15,… 200 km of the overhead transmission line simulated using PSCAD ™ / EMTDC ™ and 50 number phase a-ground faults were created at 0.2, 0.4,… 10 km of the underground cable transmission line. In Table 1, the performance results of the methods used in the Regression are given by simulating the short circuit faults that occur in the overhead and underground cable transmission lines. In the Table 2 and Table 3, the estimated fault locations and percentage error value are given as a result of the short circuit occurring in the overhead transmission line and underground cable line, respectively.

Table 1. Regression Methods Used and Performance Results

| Method | | RMSE Overhead line | RMSE Underground cable line |
|---|---|---|---|
| Linear Regression | Linear | 0.03921 | 0.010085 |
| | Interactions Linear | 0.058815 | 0.0064932 |
| | Robust Linear | 0.017652 | 0.010119 |
| | Stepwise Linear | 0.049872 | 0.0060709 |
| Tree | Fine Tree | 0.077738 | 0.061107 |
| | Medium Tree | 0.13258 | 0.12531 |
| | Coarse Tree | 0.23839 | 0.24734 |
| SVM | Linear SVM | 0.030605 | 0.016424 |
| | Quadratic SVM | 0.046394 | 0.020054 |
| | Cubic SVM | 0.24948 | 0.03174 |
| | Fine Gaussian SVM | 0.12734 | 0.071415 |
| | Medium Gaussian SVM | 0.067588 | 0.022517 |
| | Coarse Gaussian SVM | 0.069286 | 0.034763 |
| Ensemble Trees | Boosted Trees | 0.049497 | 0.046086 |
| | Bagged Trees | 0.088249 | 0.058129 |
| Gaussian Process Regression | Spuared Exponential GPR | 0.02553 | 0.0092204 |
| | Matern 5/2 GPR | 0.025624 | 0.0087294 |
| | Exponential GPR | 0.024979 | 0.0093471 |
| | Rational Quadratic GPR | 0.025474 | 0.0094911 |

Table 2. Estimated (km) Fault Locations and Percentage Error Value As A Result of Short Circuit Occurring in Overhead Transmission Line

| Method | Actual Fault Location (km) | Estimated Fault Location (km) | Error Value (%) |
|---|---|---|---|
| Linear Regression Robust Linear | 20 | 18.47938 | 0.765 |
| | 45 | 46.51063 | 0.755 |
| | 70 | 71.05625 | 0.525 |
| | 95 | 95.45563 | 0.225 |
| | 120 | 119.5869 | 0.21 |
| | 145 | 143.7181 | 0.645 |
| | 170 | 172.4563 | 1.225 |
| | 195 | 194.5644 | 0.22 |

Table 3. Estimated (km) Fault Locations and Percentage Error Value As A Result Of Short Circuit Occurring in Underground Cable Transmission Line

| Method | Actual Fault Location (km) | Estimated Fault Location (km) | Error Value (%) |
|---|---|---|---|
| Linear Regression Stepwise Linear | 0.8 | 0.835775 | 0.357 |
| | 1.8 | 1.707975 | 0.921 |
| | 2.8 | 2.687975 | 1.121 |
| | 3.8 | 3.7623 | 0.377 |
| | 4.8 | 4.817025 | 0.17 |
| | 5.8 | 5.7321 | 0.679 |
| | 6.8 | 6.720675 | 0.794 |
| | 7.8 | 7.757025 | 0.43 |
| | 8.8 | 8.66965 | 1.304 |
| | 9.8 | 9.755 | 0.45 |

## 4. Conclusions and Recommendations

In this study, fault location estimation has been made using image processing methods and Regression methods for short circuit faults occurring in mixed transmission lines. Considering the methods and training errors used as a result of the studies carried out on the mixed transmission line, it was seen that the best training model was Linear Regression in the overhead line section and Stepwise Linear in the underground cable line section. According to the test results, the highest error value is 1.225% in the estimated locations in the overhead line section and 1.304% in the underground cable line part. In the study, the fault location estimations can determine whether the fault is in the overhead line or underground cable line part in case of short circuit faults occurring in the mixed transmission line and close to the actual fault location.

As a result, in long mixed transmission lines, in case of short circuit failure, the mixed transmission line does not affect fault location estimation studies and high predictive values are shown in the tables.

The study shows that using image processing in electrical power systems can be developed and used in different ways. In this study, image processing techniques and regression methods were used for fault location estimates. In further studies, different image processing techniques can be applied, and fault location predictions can be increased by using different artificial intelligence algorithms such as ANN or Deep learning.